\documentclass{elsart}
\usepackage{graphicx}
\usepackage{amssymb}
\usepackage{t1enc} 
\usepackage[]{natbib}
\usepackage{setspace}

\newcommand{\be}{\begin{equation}}
\newcommand{\ee}{\end{equation}}
\newcommand{\ba}{\begin{array}}
\newcommand{\ea}{\end{array}}

\newcounter{ale}

\newcommand{\abc}{\item[\alph{ale})]\stepcounter{ale}}

\newenvironment{liste}{\begin{itemize}}{\end{itemize}}
\newcommand{\aliste}{\begin{liste} \setcounter{ale}{1}}
\newcommand{\zliste}{\end{liste}}

\newenvironment{abcliste}{\aliste}{\zliste}

\begin{document}

\begin{frontmatter}



\title{Predicting the asymmetric response of a genetic switch to noise}



\author{Anna~Ochab-Marcinek \corauthref{cor1}}
\ead{ochab@th.if.uj.edu.pl}
\corauth[cor1]{tel.:~(+4812)~663~5833, fax:~(+4812)~633~4079}

\address{Department of Statistical Physics,
	   M. Smoluchowski Institute of Physics, Jagellonian University
	   ul. Reymonta 4,
	   Krak\'ow, Poland}

\begin{abstract}

We present a simple analytical tool which gives an approximate insight into the stationary behavior of nonlinear systems undergoing the influence of a weak and rapid noise from one dominating source, e.g. the kinetic equations describing a genetic switch with the concentration of one substrate fluctuating around a constant mean. The proposed method allows for predicting the asymmetric response of the genetic switch to noise, arising from the noise-induced shift of stationary states. The method has been tested on an example model of the \textit{lac} operon regulatory network: a reduced Yildirim-Mackey model with fluctuating extracellular lactose concentration. We calculate analytically the shift of the system's stationary states in the presence of noise. The results of the analytical calculation are in excellent agreement with the results of numerical simulation of the noisy system. The simulation results suggest that the structure of the kinetics of the underlying biochemical reactions protects the bistability of the lactose utilization mechanism from environmental fluctuations. We show that, in the consequence of the noise-induced shift of stationary states, the presence of fluctuations stabilizes the behavior of the system in a selective way: Although the extrinsic noise facilitates, to some extent, switching off the lactose metabolism, the same noise prevents it from switching on.

\end{abstract}

\end{frontmatter}

\clearpage
\section{Introduction}

Relatively simple biochemical systems regulated at the level of gene expression are capable of a complex dynamic behavior due to their intrinsic nonlinearity. The nonlinear kinetics of the biochemical regulation may result in various patterns of behavior, among which bistability is an extremely interesting one as a source of a switch-like behavior, a common strategy used by biochemical and cellular systems to turn a graded signal into an all-or-nothing response. Another important feature associated with the bistability is hysteresis: in order to switch the system from one steady state to another, the input signal
must surpass a given threshold. To switch back, the input signal must be decreased below
another (smaller) threshold. This permits a discontinuous evolution of the system along different possible pathways, which may provide the system with an epigenetic (nongenetic) memory \citep{Laurent_Kellershohn,Casadesus_DAri,Ferrell}.

Recently, growing attention has been focused on the study of stochastic aspects of gene regulation \citep{Austin,Elowitz,Paulsson,Rosenfeld,Swain_Elowitz,Tsimring}. Fluctuations in a gene network are generally divided into 'intrinsic' and 'extrinsic'. This distinction depends on the point of view: we consider the low copy number fluctuations in a single reaction (or set of reactions) under study as the 'intrinsic noise', whereas the 'extrinsic noise' is connected with all the remaining, external processes which are not taken into consideration in detail. The extrinsic noise can originate from low copy number fluctuations in the reactions that are external with respect to the set of processes studied, as well as from other stochastically varying, unknown factors affecting our system. 

Reliable functioning of a cell may, on the one hand, require genetic networks to suppress or to be robust to fluctuations \citep{Tabaka,Elowitz,Becskei,Alon}. On the other hand, noise offers the opportunity to generate a switch-like behavior \citep{Ozbudak} and a long-term heterogeneity in a clonal population \citep{Elowitz}. The presence of fluctuations in nonlinear systems such as genetic networks may induce spontaneous switching between stationary states, emergence of new stationary states and disappearance of the existing ones \citep{Horsthemke_Lefever}.

In this paper we present a simple analytical tool which gives an approximate insight into the stationary behavior of  systems undergoing the influence of a weak and rapid noise from one nonlinear source, e.g. kinetic equations where the fluctuations of the concentration of one particular substrate dominate, and where that concentration enters into the equations in a nonlinear function. The proposed method of mean noise expansion allows for predicting the noise-induced shift of stationary states which, in case of bistable systems, causes the asymmetric response to fluctuations. This asymmetry may, for example, facilitate switching off a genetic switch but prevent it from switching on. The occurrence of such an effect can give rise to the question of a possible optimization of the genetic switch for functioning in a noisy environment.

We have tested our method on an example model of the \textit{lac} operon. The lactose regulation system in the \textit{Escherichia coli} bacteria is one of the most extensively studied examples of a biological switch: it allows for the maintenance of differences in the phenotype despite the absence of genetic and environmental differences. Monod and Pappenheimer \citep{Monod_Pappenheimer} discovered the effect of population heterogeneity on the level of an entire bacterial population, whereas Novick and Weiner \citep{Novick_Weiner} identified the same effect on the level of individual \textit{E. coli} cells: in the same external  conditions, the bacteria were either able or unable to metabolize lactose. The studies of the expression of $\beta-$galactosidase (the enzyme which breaks down the lactose into a simpler sugar) \citep{Novick_Weiner,Cohn_Horibata} and of the direct \textit{lac} gene transcription activity at the cellular level \citep{Ozbudak} show that the cells can be in one of two discrete states: either fully induced, with the transcription (and, consequently, enzyme) levels reaching a maximum, or uninduced, with negligible transcription and enzyme levels. The induction may be triggered by applying even a quite short stimulus: a temporary increase in the extracellular lactose level.

Different nonlinear dynamical models of the underlying chemical kinetics were proposed to explain the origins of the switch-like behavior of the lactose utilization network \citep{Laurent_Kellershohn,Ozbudak}. This direction of research has led to a more detailed model proposed by Yildirim and Mackey \citep{Yildirim_Mackey}. It explicitly incorporates all the relevant biochemical processes along with experimentally motivated kinetic constants, and, tested on empirical data, displays a good agreement with experiments. According to this model, the switch-like behavior of the lactose operon results from the bistability of the kinetic equations.

Since the changes of the extracellular lactose concentration are the primary factor which controls the induction and uninduction of the lactose metabolism in \textit{E. coli}, we focus our attention on this, completely external, process influencing the \textit{lac} operon system. Within the example model based on the Yildirim-Mackey framework, we analyze how the weak and rapid Gaussian fluctuations in the extracellular lactose concentration (and their different intensity) affect the \textit{lac} gene expression. We do not take into account the intrinsic fluctuations (modeling their effects deserves a separate study) but our analysis may be the first step to the interpretation of the experimental measurements of stochasticity in \textit{lac} operon expression, in terms of the discrimination between the effects of the intrinsic and extrinsic noises, which is itself a challenging task. It is worth noting that \cite{Elowitz} have shown that in systems consisting of several reactions (in particular, also in the \textit{lac} operon system in \textit{E. coli}) the extrinsic noise often gives a much stronger contribution to the gene expression than the intrinsic fluctuations.

Using the proposed method of mean noise expansion, we analytically calculate the noise-induced shift of the stationary states of the model, which gives rise to the asymmetric response of the system to fluctuations: the effective stabilization of the uninduced state and the destabilization of the induced state. We show that the results of the analytical calculation are in excellent agreement with the mean stationary states obtained from the numerical simulation of the noisy system. We also show the consequences of that shift: Varying the noise intensity, we measure mean times of the transition between the uninduced and induced states. In this way we check when the system becomes resistant to the fluctuations and when, on the contrary, the fluctuations facilitate the switching between those states. 

The paper is organized as follows: in Section~\ref{sec:theory} we present the analytical method of calculation of the noise-induced shift in stationary states of a system. In Section~\ref{sec:model} the example model of the \textit{lac} operon is described. Section~\ref{sec:results} presents the results: the application of the analytical method compared to the numerical results (Subsection~\ref{subsec:an}), and changes in the mean times of the transition between the uninduced and induced states being the consequence of the noise-induced shift (Subsections~\ref{subsec:num_iu} and~\ref{subsec:num_ui}).

\section{Theory} \label{sec:theory}
The general method of treatment of dynamical systems undergoing the influence of a weak and rapid noise from one nonlinear source, which we present below, can be applied, for example, to the models of genetic regulatory networks. A genetic switch should be described by the equations of chemical kinetics, i.e. neglecting all sources of noise, except one: a fluctuating concentration of a substrate which enters into the equations as a parameter. The concentration of that substrate should fluctuate weakly but rapidly around a constant mean.\\

Assume that:
\begin{abcliste}
\abc{ The system is described by a set of stochastic differential equations:
\be \label{eq:system_general}
\frac{d\mathbf{X}}{dt} = \mathbf{F}(\mathbf{X},h(P_t)).
\ee
$\mathbf{X}$ can denote here the vector of concentrations of reactants. $P_t$ is a parameter, for example, a concentration of a substrate, which does not depend on $\mathbf{X}$.}
\abc{$P_t$ is a stochastic process, fluctuating in time $t$ around the mean $\bar P$ and having a constant variance $\sigma^2 \ll 1$ (weak fluctuations).}
\abc{The fluctuations of $P_t$ are rapid enough not to correlate with the time scales of the processes described by the Sys.~(\ref{eq:system_general}). The characteristic time $\tau_{\mathrm{sys}}$ of the system is given by $\frac{1}{|\mathrm{Re} \lambda|}$, where $\lambda$ is the greatest eigenvalue of the Jacobian of (\ref{eq:system_general}) (a standard linearization procedure). The characteristic time scale of the process $P_t$ is determined by its correlation time $\tau$. Therefore, $\tau \ll \tau_{\mathrm{sys}}$.}
\abc{$P_t$ enters into the system ~(\ref{eq:system_general}) only in the function $h(P_t)$ only.}

\abc{The deterministic system
\be \label{eq:system_general_determ}
\frac{d\mathbf{X}}{dt} = \mathbf{F}(\mathbf{X},h(\bar P))
\ee
with a constant parameter $\bar P$ equal to the mean of $P_t$ has steady states $\mathbf{X}^*(\bar P)$.}
\end{abcliste}

If b) and c) are fulfilled, we can assume that the trajectories of the stochastic system~(\ref{eq:system_general}) fluctuate weakly and rapidly around a certain constant average $\langle \mathbf{X}(P_t) \rangle$. This means that the behavior of the system is quasi-stationary, i.e. even if the probability density of $\mathbf{X}(P_t)$ has more than one maxima, the transitions between them are very unlikely. Therefore we will consider only the trajectories which fluctuate around one of the maxima: $\langle \mathbf{X}(P_t) \rangle$ will be then the position of that maximum. Since the fluctuations are weak, the maxima of the probability density of $\mathbf{X}(P_t)$ are close to the steady states of the deterministic system (\ref{eq:system_general_determ}). 

The response of the system to noise in the parameter $P$ will be a shift of the mean, around which the trajectories of the stochastic system fluctuate,  by a small value of $\Delta$ with respect to the corresponding steady states of the deterministic system:
\be \label{eq:x}
\langle \mathbf{X}(P_t) \rangle = \mathbf{X}^*(\bar P + \Delta).
\ee

In order to find $\Delta$, we take the noisy trajectories of (\ref{eq:system_general}) at a certain time $t$ and average them over the ensamble of realizations:
\be \label{eq:avg}
0 = \langle \mathbf{F}(\mathbf{X}(P_t),h(P_t)) \rangle
\ee

At weak and rapid fluctuations, we can assume that $\mathbf{X}$ (indirectly depending on $P_t$) experiences only an averaged contribution from the noise, so it can be replaced by the constant $\langle \mathbf{X}(P_t) \rangle$, which depends on the mean $\bar P$, but also on other parameters of the process $P_t$. For example, if $P_t$ is Gaussian, i.e. fully characterized by its mean and variance, $\langle \mathbf{X}(P_t) \rangle = const(\bar P, \sigma^2)$ (it depends on the mean value of the parameter and on the strength of its fluctuation). Then, the only term in (\ref{eq:avg}) which directly depends on the fluctuating values of $P_t$ is $h(P_t)$. The averaging in (\ref{eq:avg}) can be thus separated: 
\be \label{eq:avg_h}
0 = \mathbf{F}(\langle \mathbf{X}(P_t) \rangle,\ \langle h(P_t) \rangle).
\ee
At a given time $t$, $P_t = \bar P + \delta P$ for each trajectory. Assuming that the deviations from the mean are small, we make a Taylor expansion of $h(P_t)$ around the mean $\bar P$ up to the second order:
\be \label{eq:h_taylor}
\langle h(P_t) \rangle = \langle h(\bar P + \delta P) \rangle
= h(\bar P) + h'(\bar P)\langle \delta P \rangle + \frac{1}{2}h''(\bar P)\langle \delta P^2 \rangle + ...
\ee
The average deviation from the mean $\langle \delta P \rangle=0$. The mean square deviation is equal to the variance: $\langle \delta P^2 \rangle=\sigma^2$. (If $P_t$ is a Gaussian process, then also the third-order term $\langle \delta P^3 \rangle=0$). The Eq.~(\ref{eq:h_taylor}) is now approximated by:
\be \label{eq:avg_h_appr}
\langle h(P_t) \rangle = \ h(\bar P) + \frac{1}{2}h''(\bar P)\sigma^2.
\ee
The deterministic steady states in (\ref{eq:x}) are given by the equation:
\be \label{eq:determ}
0 = \mathbf{F}(\mathbf{X^*}(\bar P+\Delta),\ h(\bar P+\Delta)),
\ee
whereas (\ref{eq:avg_h}) is now
\be \label{eq:stoch}
0 = \mathbf{F}(\langle \mathbf{X}(P_t) \rangle,\ h(\bar P) + \frac{1}{2}h''(\bar P)\sigma^2).
\ee
Using (\ref{eq:x}), we can compare (\ref{eq:determ}) and (\ref{eq:stoch}). The approximate value of the shift $\Delta$ can be therefore calculated from
\be \label{eq:heq}
h(\bar P + \Delta) =h(\bar P)+ \frac{1}{2} h''(\bar P) \sigma^2.
\ee
For the weak noise, the shift should be small, so it can be further approximated by the Taylor expansion of $h(\bar P + \Delta)$ up to the first order in $\Delta$. The Eq.~(\ref{eq:x}) then takes the form:
\be \label{eq:happrox}
\langle \mathbf{X}(P_t) \rangle= \mathbf{X}^*(\bar P + \frac{h''(\bar P)}{2 h'(\bar P)}\sigma^2),
\ee
if $h'(\bar P) \neq 0$.

The approximation lies in the fact that we attribute the shift of the stationary states to the functional dependence of $h(P_t)$ only. In some cases, the multiplicative dependence on the system's variables, $h(P_t)f(\mathbf{X}(P_t))$, may be also important. Here we neglect it assuming that the noise is weak and rapid, so that the system variables can be treated as constant. In case our method does not show the noise-induced shift of stationary states (for example when $h''\equiv0$, i.e. the noisy parameter enters linearly into $\mathbf{F}$), it may still turn out that a noise-induced shift is present, namely because the multiplicative dependence of the noise becomes important \citep{Horsthemke_Lefever}.

\section{Example: lactose switch model} \label{sec:model}

Below we present an example model of \textit{lac} operon regulatory network to which our method will be applied.

 \begin{figure}[t]
    \begin{center}
      \includegraphics*[width=5in]{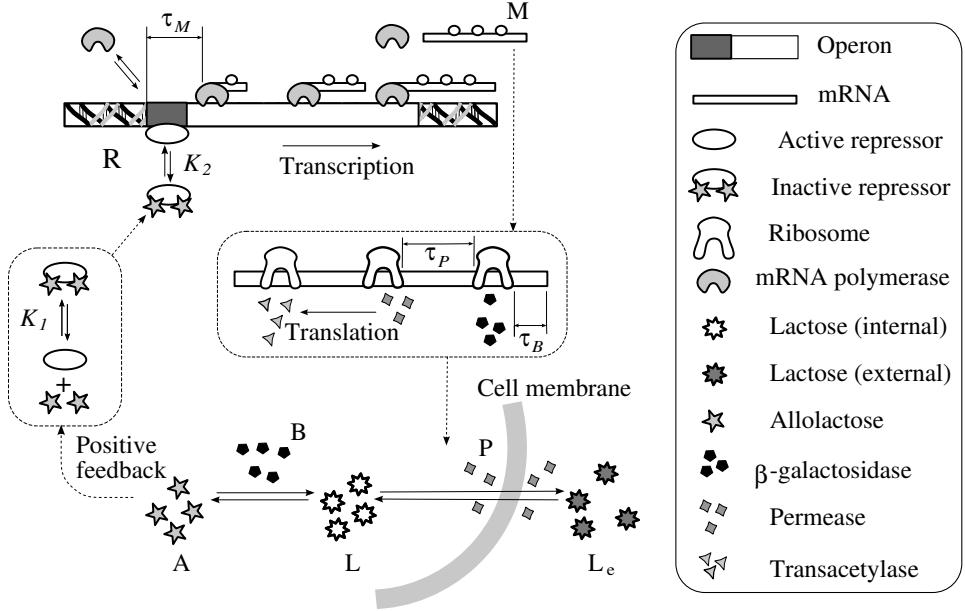}
     \caption{\label{fig:scheme} Schematic representation of the lactose operon regulatory system.
}
   \end{center}
\end{figure}

\subsection{The regulation mechanism of the \textit{lac} operon} \label{subsec:scheme}

The scheme of the feedback regulation mechanism of the induction in the \textit{lac} operon in \textit{\textit{E. coli}} \citep{Yildirim_Mackey} is presented in Fig.~\ref{fig:scheme}: The extracellular lactose ($L_e$) is transported through the cell membrane by the enzyme permease ($P$). The intracellular lactose ($L$) is then broken down into glucose, galactose, and allolactose ($A$) by the enzyme $\beta$-galactosidase ($B$). By a positive feedback loop, the presence of allolactose enables the production of permease and $\beta$-galactosidase enzymes.

The \textit{lac} operon consists of three genes, $lac$Z, $lac$Y, and $lac$A preceded by a promoter/operator region. The $lac$Z gene encodes for the mRNA responsible for the production of $\beta$-galactosidase, whereas the $lac$Y gene produces the mRNA for permease. (The third gene, $lac$A, producing thiogalactoside transacetylase, does not play a role in the \textit{lac} operon regulation \citep{Beckwidth}.) The transcription of this part of DNA is controlled by a repressor protein ($R$). If the inducer (allolactose) is absent, the repressor binds to the operator DNA sequence and makes the transcription of genes by the RNA polymerase impossible. In the presence of inducer, a complex is formed between allolactose and the repressor that prevents the latter from binding to the operator region. The RNA polymerase is then able to initiate the transcription of the $lac$Z, $lac$Y, and $lac$A genes to produce mRNA ($M$). Subsequently, the mRNA is translated into the apropriate enzymes ($P$ and $B$).

\subsection{Yildirim-Mackey model} \label{subsec:ym}

Our study is based on the model of \textit{lac} operon regulation proposed by Yildirim and Mackey \citep{Yildirim_Mackey}. The model consists of five equations of chemical kinetics for the reactions described in Subsection~\ref{subsec:scheme}, involving mRNA ($M$), allolactose ($A$), lactose ($L$), $\beta$-galactosidase ($B$) and permease ($P$):
\be \label{eq:ym}
\ba{lll}
\frac{dM}{dt} & = & \alpha_M \ f_1(e^{-\mu \tau_M } \ A_{\tau_M}) + \Gamma_0 - \tilde{\gamma}_M \ M \\
\frac{dB}{dt} & = & \alpha_B \ e^{-\mu \tau_B } \ M_{\tau_B} - \tilde{\gamma}_B \ B \\
\frac{dA}{dt} & = & \alpha_A \ B \ g_1(L) - \beta_A \ B \ f_2(A) - \tilde{\gamma}_A \ A \\
\frac{dL}{dt} & = & \alpha_L \ P \ h(L_e) - \beta_L \ P \ g_2(L) - \alpha_A \ B \ g_1(L) - \tilde{\gamma}_L \ L\\
\frac{dP}{dt} & = & \alpha_P \ e^{-\mu (\tau_P + \tau_B) } \ M_{\tau_P + \tau_B}- \tilde{\gamma}_P \ P.
\ea
\ee
$f_1(A)$ describes the effector-repressor dynamics of the transcription enhanced by allolactose:
\be
f_1(A)=\frac{1+K_1 \ A^2}{K + K_1 \ A^2}
\ee 
The square term derives from the average number of allolactose molecules required to inactivate the repressor, which (according to \cite{Yildirim_Mackey,Yagil}) is approximately equal to 2. $f_2$, $g_1$, $g_2$ and $h$ are rates of the irreversible Michaelis-Menten reactions:
\be \label{eq:funcs}
\ba{lll}
f_2(A) = \frac{A}{K_A + A}, & \ \ & g_1(L) = \frac{L}{K_L + L},\\
g_2(L) = \frac{L}{K_{L1} + L}, & \ \ & h(L_e) = \frac{L_e}{K_{L_e} + L_e}
\ea
\ee
$\alpha$ and $\beta$ denote the gain and loss rates for the reactions. $K_1$ is the equilibrium constant for the repressor-allolactose reaction. $K_2$ is the equilibrium constant for the operator-repressor reaction, $K=1+K_2 R_{tot}$, and $R_{tot}$ is the total amount of the repressor. $\tau$ represents the delays associated with the finite time required to complete the transcription ($\tau_M$) and the translation ($\tau_P$ and $\tau_B$). For example, $A_{\tau_M} \equiv A(t-\tau_M)$. The $\tilde{\gamma} = \gamma + \mu$ are the coefficients for the terms representing decay of species due to chemical degradation ($\gamma$) and dilution ($\mu$). The exponential factors take into account the dilution of mRNA due to cell growth.  Even if allolactose is totally absent, on occasion repressor will transiently not be bound to the operator and RNA polymerase will initiate transcription. Although the mRNA production rate $dM / dt$ would be then nonzero (a leakage transcription), it is necessary to add an empirical constant $\Gamma_0$ to the model to obtain a leakage rate that agrees with experimental values \citep{Yildirim_Mackey}. A more detailed derivation of the above equations, an elaborate estimation of the parameter values, as well as the results of testing the model on experimental data can be found in \citep{Yildirim_Mackey}. The model reproduces the bistable behavior of the lactose operon for realistic extracellular lactose concentrations.
%
\subsection{Reduced model}\label{app:reduced}
In this paper, we study a reduced model derived from the full Yildirim-Mackey equations. The simplified model is less time-consuming for numerical simulations (Subs.~\ref{subsec:num_ui} and \ref{subsec:num_iu}) and easier for analytical treatment (Subs.~\ref{subsec:an}) but, at the same time, it captures all the important features of the full model. 

Our model consists of three equations of kinetics for mRNA ($M$), allolactose ($A$) and lactose ($L$) concentrations in the bacterial cell:
\be \label{eq:mysys}
\ba{lll}
\frac{dM}{dt} & = & \alpha_M \ \frac{1+K_1 \ A^2}{1+K_2 R_{tot} + K_1 \ A^2} + \Gamma_0 - \tilde{\gamma}_M \ M \\
&&\\
\frac{dA}{dt} & = & k_B \ M \left(\alpha_A \ \frac{L}{K_L + L} - \beta_A \ \frac{A}{K_A + A}\right) - \tilde{\gamma}_A \ A \\
&&\\
\frac{dL}{dt} & = & k_P \ M \left(\alpha_L \ \frac{L_e}{K_{L_e} + L_e} - \beta_L \ \frac{L}{K_{L1} + L}\right) - \alpha_A \ k_B \ M \ \frac{L}{K_L + L} - \tilde{\gamma}_L \ L \\
\ea
\ee
The model has been obtained from Eqs.~(\ref{eq:ym}) by the assumption that: i) The enzyme concentration is proportional to the mRNA concentration. The assumption is based on the fact that the amount of proteins observed in procariotic cells is in general proportional to the amount of their transcripts (which is, in particular, the case for the \textit{lac} operon in \textit{E. coli}) \citep{Mathews_vHolde_Ahern}.

\be \label{eq:enzymes}
\ba{lll}
B & = & \frac{\alpha_B \ e^{-\mu \tau_B}}{\tilde{\gamma}_B} M = k_B M\\
P & = & \frac{\alpha_P \ e^{-\mu (\tau_P+\tau_B)}}{\tilde{\gamma}_P} M = k_P M
\ea
\ee
ii) The time delays are small enough to fullfill:
$e^{-\mu \tau_x} \simeq 1$, and iii) $X(t+\tau_x) \simeq X(t)$.
 The values of parameters estimated by \citep{Yildirim_Mackey} (see Table~\ref{tab:params}) allow for such assumptions. Assumption iii) is very well fullfilled in the neighborhood of stationary points, but also the differences between $X(t)$ and $X(t+\tau_x)$ in transient regions are not large compared to the distance between stationary points (see the trajectories in Figs.~\ref{fig:ind_unind},~\ref{fig:unind_ind}).

\begin{figure}[t]
   \begin{center}
	\includegraphics*[width=5in]{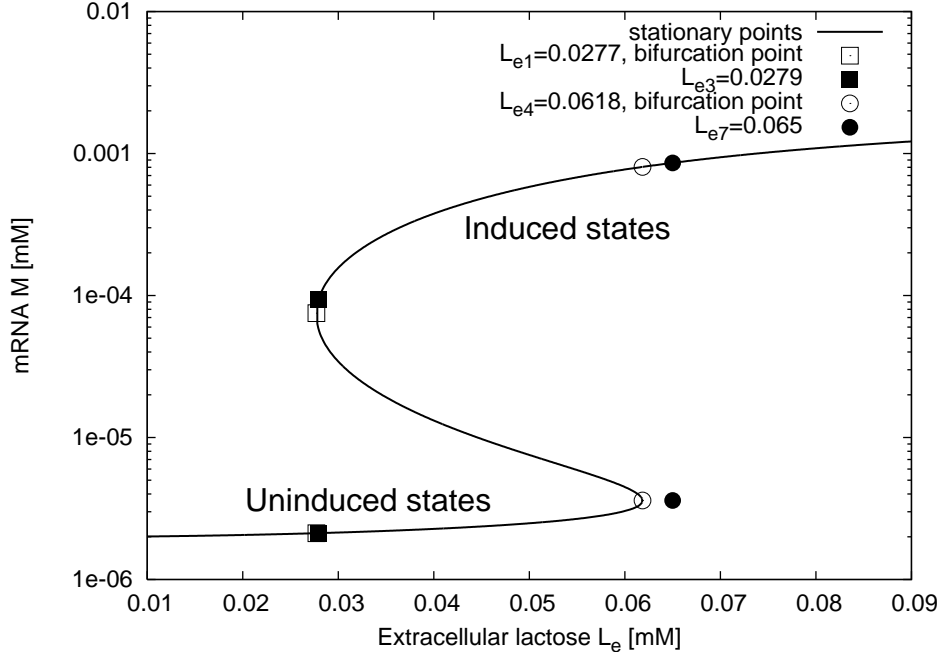}
	\caption{\label{fig:steady_statesM} Steady states of the system (for mRNA concentration). Marked are the initial and final points of the simulations described in Section~\ref{sec:results}.}
   \end{center}
\end{figure}

The stability properties of the model reproduce very well those of the full Yildirim-Mackey model (see Fig.~\ref{fig:steady_statesM} and compare with \citep{Yildirim_Mackey}): for the concentrations of the extracellular lactose
\be 
0.0277 \ \mathrm{mM} \ < L_e < 0.0618 \ \mathrm{mM},
\ee 
the system has two steady states (induced and uninduced). Assuming that the extracellular lactose concentration is constant and does not fluctuate in time, the model shown above is deterministic. If $L_e$ lies in the range of bistability, then, after a sufficiently long time, the concentrations of allolactose, lactose and mRNA reach one of the steady states. Switching between the two steady states is then impossible unless the extracellular lactose concentration changes (increasing or decreasing $L_e$ beyond the bistable region would enable a hysteretic response of the system). 

\subsection{Fluctuations in the extracellular lactose concentration} \label{sec:noise}

We investigate how the fluctuations in the extracellular lactose concentration induce the shift of stationary states of the system and spontaneous transitions between those states.
To model the fluctuations in $L_e$, we have chosen Gaussian noise, assuming that the natural environment, where wild-type bacteria remain, undergoes the influence of very many random factors. If their contributions are independent, they add together into the noise of a Gaussian distribution around a certain mean. We therefore assume that the noise in $L_e$ is Gaussian and fast with respect to other processes in the system. We model it using the Ornstein-Uhlenbeck process \citep{Gardiner}
\be \label{eq:OU}
\frac{d L_e}{dt} = - \theta (L_e - \bar{L}_e) + \gamma \xi(t)
\ee
which fluctuates in time around the mean value $\bar{L}_e$. $\xi(t)$ is a Gaussian white noise of intensity $\gamma$ and autocorrelation $\langle \xi(t) \xi(s) \rangle = \delta(t-s)$. The correlation time of the fluctuations in $L_e$, $\tau_{OU}=1 / \theta =2 \cdot 10^{-2}\ \mathrm{min} = 1.2 \ \mathrm{s}$,  has been chosen significantly larger than the time step in the numerical calculations ($\delta t=2 \cdot 10^{-3}  \  \mathrm{min}$) but smaller than the fastest time scale of the system $\tau_{\mathrm{sys}} \sim 10^{-1} \ \mathrm{min}$ (see Appendix~\ref{app:timescales}). In this way we obtain a rapidly fluctuating stochastic process which does not correlate with the time scales of the biochemical processes described by the model. Otherwise, if the noise were as slow as the system's reaction to it, the $A$, $L$, and $M$ concentrations would have enough time to adapt to the instantaneous values of $L_e$. In that case, the further analytical treatment described in this paper would be impossible. Taking into account the mobility of the \textit{E. coli} bacteria (mean velocity $\sim 30 \mu \mathrm{m / s}$, see e.g. \cite{DiLuzio}), granularity of the intestinal content and motions of intestinal villi, we may suppose that the fluctuation rapidity assumed here is realistic. The variance of the extracellular lactose fluctuations is $\gamma^2 / (2 \theta)$ \citep{Gardiner}. The value of $\gamma$ controlling the noise intensity will be varied in our simulations.

\section{Results}
\label{sec:results}

Our goal was to determine how the model of the \textit{lac} gene expression is sensitive to fluctuations in the  extracellular lactose concentration. Using the proposed method of noise expansion, we calculated analytically the noise-induced shift of the stationary states of the system~(\ref{eq:mysys}) with the external lactose concentration $L_e$ perturbed by the Gaussian noise given by the Eq.~(\ref{eq:OU}). 

In order to show the effects of the asymmetric response of the system to noise, we performed numerical simulations of the trajectories of the system (see Appendix~\ref{app:simulation} for detailed technical information). We measured the mean time of noise-driven transitions between the induced and uninduced states for different values of the mean extracellular lactose concentration $\bar{L}_e$ and with different noise intensities $\gamma$ (see Appendix~\ref{app:simulation} for the simulation details). To show the fluctuation strength on the graphs in a more intuitive way, the noise intensity unit used in the below-presented figures is the standard deviation of $L_e$, equal to $\sqrt{\gamma^2 / (2 \theta)}$.


\subsection{Analysis of noise-induced stabilization and destabilization}\label{subsec:an}


\begin{figure}[t]
   \begin{center}
	\includegraphics*[width=5in]{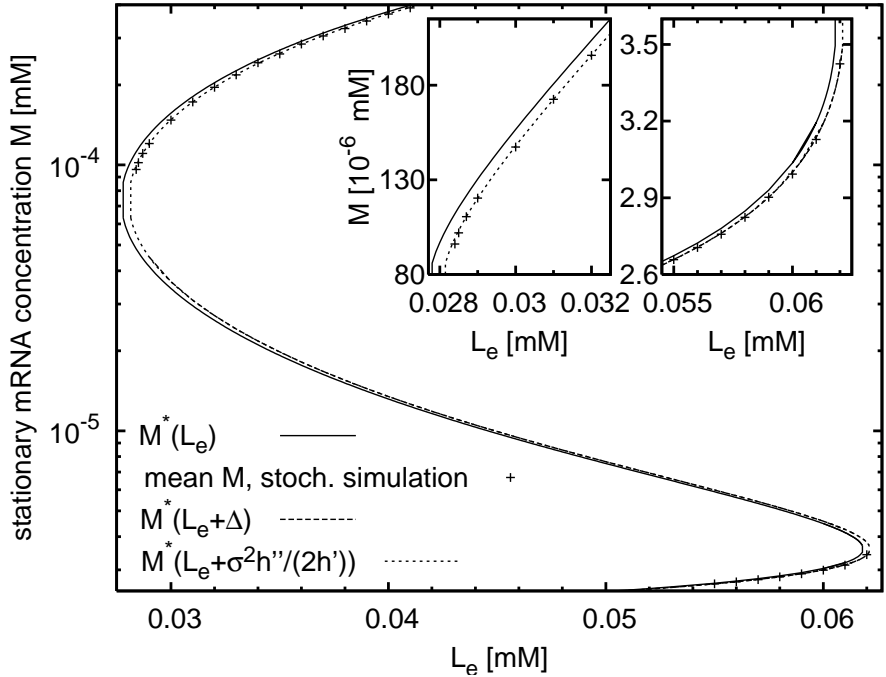}
\caption{\label{fig:analytical} Stationary states $\langle M(\bar L_e) \rangle$ for mRNA concentration at different mean concentrations of extracellular lactose, found using the approximate method of noise expansion. Under the influence of noise  ($\sigma^2$ = 0.01 mM), the stationary states shift right by $\Delta$  with respect to the deterministic stationary states $ M^*(\bar L_e)$. The dashed lines for $\Delta$ calculated from the Eq.~(\ref{eq:heq}) and for $\Delta = \frac{\gamma^2}{\theta(K_{L_e} + \bar{L}_e)}$ (Eq.~(\ref{eq:happrox})) overlap. Comparing with the results of the numerical simulations (+), one can note that the approximation is excellent.}
   \end{center}
\end{figure}

\begin{figure}[t]
   \begin{center}
	\includegraphics*[width=4.7in]{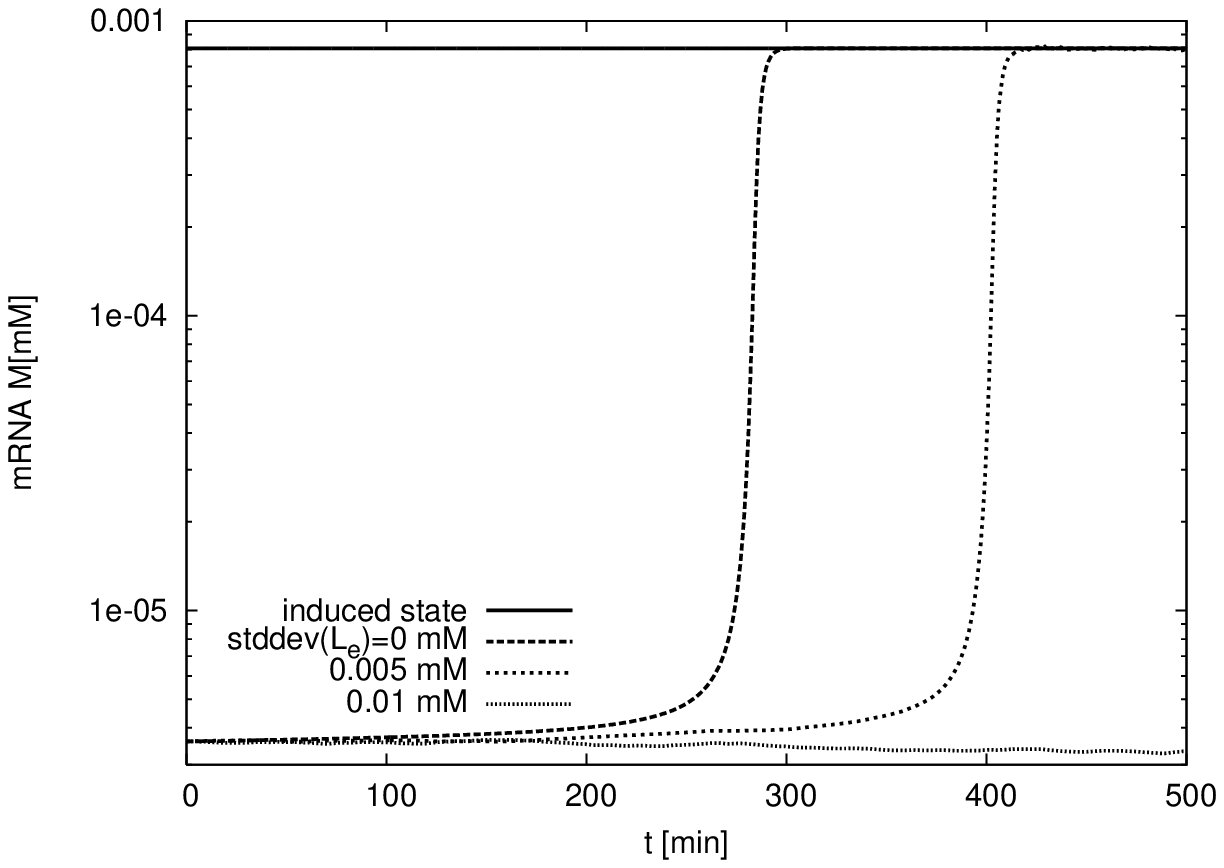}\\
	\includegraphics*[width=4.7in]{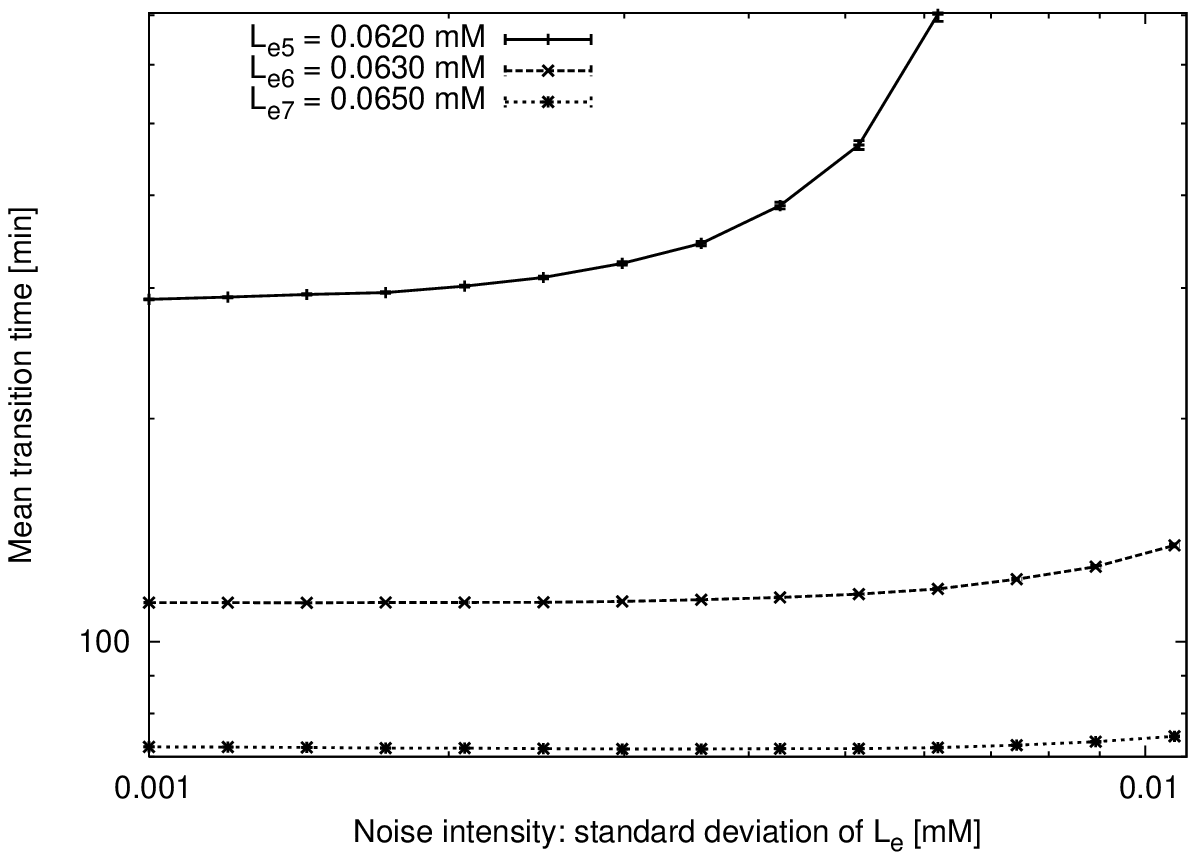}
\caption{\label{fig:unind_ind} Transition from uninduced to induced state. Simulation of changes in mRNA concentration, depending on the strength of fluctuations in extracellular lactose concentration. Top panel: Example realizations  (single trajectories) for mean extracellular lactose concentration $\bar{L}_{e5}=0.0620$ (beyond the bistable region) and different noise intensities. We have chosen the starting points same as the coordinates of the steady state at the bifurcation point $\bar{L}_{e4}=0.0618$ (see Fig.~\ref{fig:steady_statesM}). For $\mathrm{stddev}(L_e) = 0.010$, the fluctuations make the induction impossible (also, the mean mRNA concentration shifts down). In other cases, the noise causes a delay of the induction. Bottom panel: Mean times of the transition beyond the bistable region for $\bar{L}_{e5}=0.0620$, $\bar{L}_{e6}=0.0630$ and $\bar{L}_{e7}=0.0650$, depending on the noise intensity. We have chosen the starting points same as the coordinates of the steady state at the bifurcation point, and the final points as the deterministic steady states (see Fig.~\ref{fig:steady_statesM}). Here, the stronger the fluctuations are, the longer is the mean transition time, i.e. the more difficult is the induction.}
   \end{center}
\end{figure}

To analytically predict the behavior of the steady states of the system under the influence of noise, we use the approximate method of mean noise expansion presented in Sec.~\ref{sec:theory}. Treating the instantaneous value of $L_e(t)$ as a small perturbation from the mean value, we expand the noise in $L_e$ around its mean and average the equations of kinetics over an ensemble of possible trajectories, to get the mean stationary concentrations of $A$, $L$, $M$ in the noisy system. The fluctuating parameter $L_e$ enters into the equations (\ref{eq:mysys}) in the Michaelis-Menten form $h(L_e) = \frac{L_e}{K_{L_e} + L_e}$. In the Fig.~\ref{fig:analytical} we compare the noise-induced shifts of stationary states of the mRNA concentration: (i) obtained from the numerical simulations, (ii) calculated by numerical solution of the Eq.~(\ref{eq:heq}), (iii) and by the further approximation~(\ref{eq:happrox}) which here takes the form of 
\be
\Delta = \frac{h''(\bar{L}_e)}{2 h'(\bar{L}_e)}\sigma^2 = - \frac{\gamma^2}{\theta(K_{L_e} + \bar L_e)}.
\ee
All the results are in excellent agreement: positions of the stationary states in the bistable regime shift right with increasing noise intensity, which results in the destabilization of the induced steady state for small $\bar{L}_e$ and the stabilization of the uninduced steady state for large $\bar{L}_e$. The steady-state mean mRNA  concentrations for a given value of $\bar{L}_e$ decrease due to noise (compare with Fig.~\ref{fig:unind_ind}).

\subsection{Transition from the induced to uninduced state}\label{subsec:num_iu}

\begin{figure}[t]
   \begin{center}
	\includegraphics*[width=5.0in]{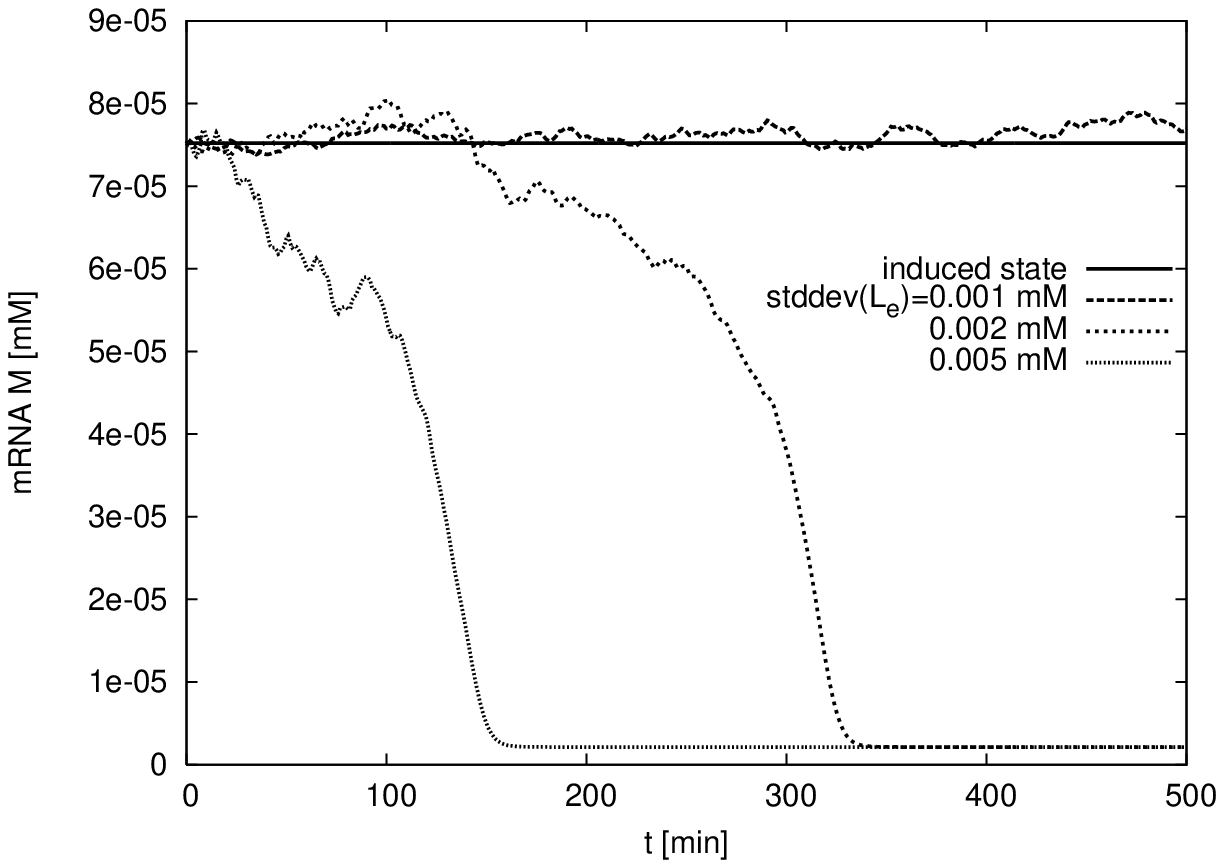}\\
	\includegraphics*[width=5.0in]{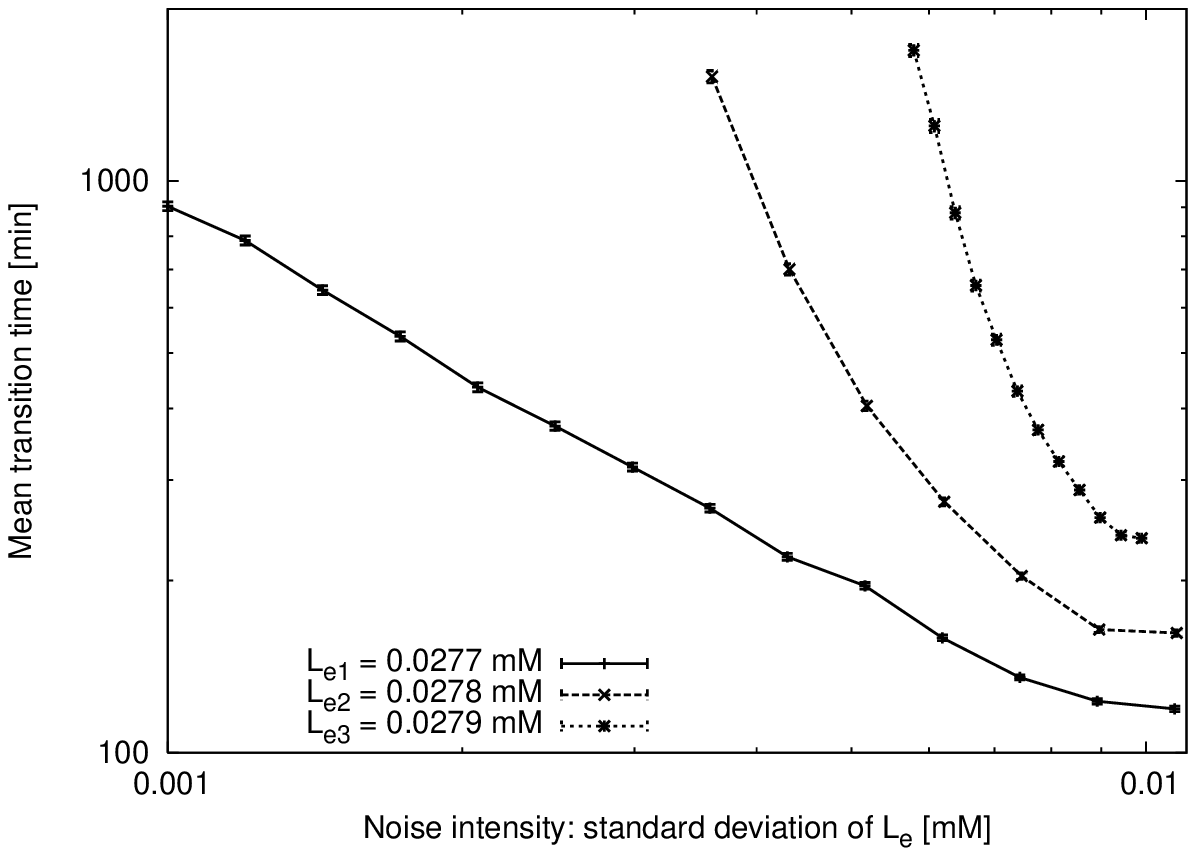}
	\caption{\label{fig:ind_unind} Transition from the induced to uninduced state. Simulation of changes in mRNA concentration, depending on the strength of fluctuations in extracellular lactose concentration (here, for clarity, measured by the standard deviation of $L_e$). Top panel: Example realizations (single trajectories) for mean extracellular lactose concentrations $\bar{L}_{e1}=0.0277$ mM (see Fig.~\ref{fig:steady_statesM}) and different noise intensities. Bottom panel: Mean times of the transition for $\bar{L}_{e1}=0.0277$, $\bar{L}_{e2}=0.0278$ and $\bar{L}_{e3}=0.0279$, depending on the noise intensity. The initial and final points for the trajectories are same as in Fig.~\ref{fig:steady_statesM}). A slight increase in the mean extracellular lactose concentration from $\bar{L}_{e1}$ to $\bar{L}_{e2}$ causes a drastic increase of the mean transition time.}
   \end{center}
\end{figure}

In order to show the consequences of the noise-induced shift of the stationary states, we performed a series of simulations in which the system started in the induced state. The initial state of the system was given by the deterministic (i.e. calculated in the absence of noise) stationary concentrations of $A$, $L$ and $M$ at the induced state for a given mean extracellular lactose level $\bar{L}_e$ (see Fig.~\ref{fig:steady_statesM} and Table~\ref{tab:points}). The results (Fig.~\ref{fig:ind_unind}) show that the system is generally resistant to the fluctuations in the extracellular lactose concentration. However, the noise slightly destabilizes the system, i.e. it enables switching from the induced to uninduced state. Such transitions are only possible for $\bar{L}_e$ very close to the bifurcation point $L_e=0.0277$ (the beginning of the bistable region).  A small increase in the mean extracellular lactose concentration causes a drastic increase of the mean transition time. For example, after changing the mean extracellular lactose concentration from $\bar{L}_{e1}=0.0277$ mM to $\bar{L}_{e2} = 0.0279$ mM, the noise intensity (measured by variance $\gamma^2 / (2 \theta)$) must be almost one order of magnitude stronger to force the transition to occur in a similar time. Within the simulation time (1000 min) no backward transitions have been recorded. The graph of mean transition times vs. noise intensity shows that the stronger the fluctuations are, the shorter is the mean transition time i.e. the easier is the uninduction. As we will see in the next subsection, the noise-driven induction depends on the noise strength in quite a different manner.

 \clearpage
\clearpage


 \clearpage
\clearpage

\subsection{Transition from the uninduced to induced state}\label{subsec:num_ui}

Next, we performed simulations which started from the uninduced state (See Fig.~\ref{fig:steady_statesM} and Table~\ref{tab:points}). Here we observe that, on the contrary to the previous simulations, the fluctuations of the extracellular lactose concentration stabilize the uninduced state (see Fig.~\ref{fig:unind_ind}). In the bistable region (increasing $\bar{L}_e$ up to the bifurcation point $L_{e4}=0.0618$, i.e. to the end of the bistable region), even very strong fluctuations cannot force the induction. Beyond the bistable region ($\bar{L}_{e5}=0.0620$, $\bar{L}_{e6}=0.0630$, $\bar{L}_{e7}=0.0650$), we have chosen the starting points same as the coordinates of the steady state at $L_{e4}$, and final points as the deterministic induced steady states for $\bar{L}_{e5}$, $\bar{L}_{e6}$ and $\bar{L}_{e7}$.
In this case, the noise causes a delay of induction, and sufficiently strong fluctuations make the induction impossible at all, whereas the mean mRNA concentration even decreases. The dependence of the mean transition time on the noise intensity is opposite than in the case of uninduction: here, the stronger the fluctuations are, the longer is the mean transition time, i.e. the more difficult is the induction.

\section{Summary} \label{sec:summary}

We present a simple analytical tool which gives an approximate insight into the stationary behavior of nonlinear systems undergoing the influence of a weak and rapid noise from one dominating source. These can be, for example, the kinetic equations describing a genetic switch with the concentration of one substrate fluctuating around a constant mean. The method is applicable in those cases when the noise enters into the equations in a form of a nonlinear function of a parameter which fluctuates weakly and rapidly, so that the fluctuations can be assumed to contribute to the state of the system as an average (i.e. when the noise is rapid enough not to directly couple with the system's dynamics). The proposed method of mean noise expansion allows for predicting the effect of the asymmetric response to noise, which, for example, may facilitate switching off a genetic switch but prevent it from switching on. 

The method has been tested on the example model of the \textit{lac} operon regulatory network: a reduced Yildirim-Mackey model \citep{Yildirim_Mackey} subject to fluctuations in extracellular lactose concentration $L_e$. We analyzed how the fluctuations, modelled by the Gaussian noise, and their different intensities affect the \textit{lac} gene expression predicted by the model. Using the method of mean noise expansion, we calculated analytically the shift of the system's stationary states in the presence of the fluctuations. We have shown that the results of the analytical calculation are in excellent agreement with the  mean stationary states obtained from the numerical simulation of the noisy system. 

The shift of stationary states gives rise to the asymmetric response of the system to the noise: the effective stabilization of the uninduced state and the destabilization of the induced state. We show the consequences of that shift by analyzing the mean times of the numerically simulated noise-driven transitions between induced and uninduced states in the bistable region (i.e. in the range of extracellular lactose concentration $L_e$ where both induced and uninduced states are possible). The simulation results show that the system as a whole is highly resistant to fluctuations. It can be stochastically induced only for the lowest possible $L_e$, i.e. very close to the left boundary of the bistable region, and only at a high noise intensity. On the other hand, stochastic induction is impossible in the bistable region even for the highest possible $L_e$, i.e. in the closest neighborhood of the right boundary of the bistable region. Moreover, noise inhibits induction even beyond the bistable region: if fluctuations in $L_e$ are strong enough, the lactose metabolism remains uninduced, although at weaker fluctuations or in their absence it would switch to the induced state. Thus, in the result of the steady-state shift, the influence of noise on the induction and uninduction of lactose metabolism is quite opposite: whereas the spontaneous uninduction (although difficult) is facilitated by noise, the same noise causes the suppression of the induction.

The above results suggest that the bistability of the lactose utilization mechanism is ``protected'' by the structure of the kinetics of the underlying biochemical reactions: switching the lactose switch due to an extracellular fluctuation is very difficult. Moreover, the ``protection level'' is different for the induction and uninduction: the possibility of random switching on the lactose metabolism is much more strongly protected than the possibility of random switching it off. Although, on the mathematical level, such a behavior results from the structure of the kinetic equations, a question may be posed if this particular shape of kinetics can have a deeper explanation, for example whether the ``protection'' of the switch against the external fluctuations is connected with preventing an unnecessary energetic effort? This suggests a new direction of study: an analysis of the noisy system of lactose metabolism in \textit{E. coli} from the energetic point of view.

It is worth noting that the noise-induced shift of stationary states and the emergence of new stationary states in systems with multiplicative noise (such as the system under study) are the effects which may be described in terms of the stationary probability distributions. These can be obtained by solving the Fokker-Planck equation associated to the noisy equations of kinetics, treated as a multi-dimensional Langevin equation \citep{Horsthemke_Lefever, Gardiner}. Further study in this direction seems to be an interesting perspective. The effect of intrinsic fluctuations as well as the response of the Yildirim-Mackey model to slow extrinsic noise also deserves a separate study.

\section{Acknowledgements}

The author thanks prof. M{\aa}ns Ehrenberg (Department of Cell and Molecular Biology, Uppsala University), dr. Sylwia K\k{e}dracka-Krok (Department of Physical Biochemistry, Jagellonian University), and prof. Ewa Gudowska-Nowak (Department of Statistical Physics, Jagellonian University) for inspiring discussions.

\newpage
\appendix

\section{Details of the reduced Yildirim-Mackey model} \label{app:timescales}

Table \ref{tab:params} presents the parameters of the model (same as used in \citep{Yildirim_Mackey}) and the data in the Table \ref{tab:statp} shows the stationary states of the system for two example values of external lactose concentration, close to bifurcation points. Table \ref{tab:statp}
\begin{table}[h!]
\begin{center}
\begin{tabular}{|l|l||l|l|} \hline
$\Gamma_0$ & $7.25 \times 10^{-7} \ \mathrm{mM/min}$&$\mu$ & $0.0226 \ \mathrm{min}^{-1}$\\
$\alpha_A$ & $1.76 \times 10^4 \ \mathrm{min}^{-1}$ &$\tau_B$ & $2.0 \ \mathrm{min}$\\
$\alpha_B$ & $1.66 \times 10^{-2} \ \mathrm{min}^{-1}$ &$\tau_M$ & $0.1 \ \mathrm{min}$\\
$\alpha_L$ & $2.88 \times 10^3 \ \mathrm{min}^{-1}$ & $\tau_P$& $0.83 \ \mathrm{min}$\\
$\alpha_M$ & $9.97 \times 10^{-4} \ \mathrm{mM/min}$ & $K$& $7.2 \times 10^3$\\
$\alpha_P$ & $10.0 \ \mathrm{min}^{-1}$& $K_1$& $2.52 \times 10^4 \ \mathrm{mM}^{-2}$\\
$\beta_A$ & $2.15 \times 10^4 \ \mathrm{min}^-1$ & $K_A$ & $1.95 \ \mathrm{mM}$\\
$\beta_L$ & $2.65 \times 10^3 \ \mathrm{min}^{-1}$ & $K_L$& $0.97 \ \mathrm{mM}$\\
$\gamma_A$ & $0.52 \ \mathrm{min}^{-1}$ & $K_{L_e}$& $0.26 \ \mathrm{mM}$\\
$\gamma_B$ & $8.33 \times 10^{-4} \ \mathrm{min}^{-1}$ & $K_{L_1}$& $1.81 \ \mathrm{mM}$ \\
$\gamma_L$ & $0.0 \ \mathrm{min}^{-1}$ & $k_B$& $0.677$\\
$\gamma_M$ & $0.411 \ \mathrm{min}^{-1}$& $k_P$& $13.94$\\
$\gamma_P$ & $0.65 \ \mathrm{min}^{-1}$& &\\
\hline
\end{tabular}
\end{center}
\vspace*{0.5cm}
\caption{\small Parameters of the model (same as used in \citep{Yildirim_Mackey}).}\label{tab:params}
\end{table}

\begin{table}[h!]
\begin{small}\begin{center}
Stationary states [mM]:\\

\begin{tabular}{|l|lll|lll|}
\hline
&\multicolumn{3}{|c|}{Uninduced} & \multicolumn{3}{c|}{Induced}\\
\hline
$L_e \mathrm{[mM]}$ & $A^*$ & $L^*$ & $M^*$ & $A^*$ & $L^*$ & $M^*$\\
\hline
$0.0278$ & $4.00 \cdot 10^{-3}$ & $9.40 \cdot10^{-2}$ & $2.12 \cdot10^{-6}$& 
 $1.04 \cdot10^{-2}$ & $0.129$ & $7.52 \cdot10^{-5}$ \\
\hline
$0.0610$ &$1.22 \cdot10^{-2}$ & $0.216$ & $3.19 \cdot10^{-6}$&
$0.386$ & $0.280$ & $7.90 \cdot10^{-4}$\\
\hline

\end{tabular}
\end{center}\end{small}\vspace*{0.5cm} \caption{\label{tab:statp}Stationary states of the system for two example values of external lactose concentration, close to the bifurcation points. Table \ref{tab:times} presents the characteristic times of the system calculated for the stationary points given in Table~\ref{tab:statp}.}
\end{table}


 
\begin{table}[h!]
\begin{small}\begin{center}
Characteristic times [min]:\\
\begin{tabular}{|lll|lll|lll|lll|}
\hline
\multicolumn{6}{|c|}{$L_e=0.0278 \mathrm{mM}$} & \multicolumn{6}{c|}{$L_e=0.0610 \mathrm{mM}$}\\
\hline
\multicolumn{3}{|c}{Uninduced} & \multicolumn{3}{c|}{Induced} &\multicolumn{3}{c}{Uninduced} & \multicolumn{3}{c|}{Induced}\\
\hline
$\tau_1$ & $\tau_2$ & $\tau_3$ & $\tau_1$ & $\tau_2$ & $\tau_3$&$\tau_1$ & $\tau_2$ & $\tau_3$ & $\tau_1$ & $\tau_2$ & $\tau_3$\\
\hline
 1.4 & 3.1 & 13 & 43 & 0.64 & 0.42 & $1.1$ & $6.4$ & $41$ & $0.21$ & $2.7$& $0.055$\\
\hline
\end{tabular}
\vspace{1ex}
\end{center} \end{small} \vspace*{0.5cm}
\caption{Characteristic times of the system calculated for the stationary points given in Table~\ref{tab:statp}\label{tab:times}}
\end{table}

\section{Simulation details} \label{app:simulation}
We performed the numerical simulations of the trajectories of the system described by the Eqs.~(\ref{eq:mysys}) with the lactose concentration $L_e$ perturbed by the Ornstein-Uhlenbeck noise given by the Eq.~(\ref{eq:OU}). The equations were solved numerically using the Euler scheme \citep{nr,Mannella} with the timestep $\delta t=2\cdot 10^{-3}$. The simulations were run starting from one of the deterministic steady states for a given $\bar{L}_e$. The time was measured until the stochastic system's trajectory gets into a close neighborhood (of a radius $D = \sqrt{\delta A^2 + \delta L^2 + \delta M^2} = 0.005$ mM) of the other deterministic stationary state. The initial and final points of the simulations are presented in Table~\ref{tab:points}. The simulation was monitored to prevent fluctuating concentrations become negative but no such event occured during the simulations. The number of simulation runs was $N=1000$.

\begin{table}[h!]
\begin{small}\begin{center}
\begin{tabular}{|l|l|l|l|l|l|l|l} \hline
$\bar{L}_e [\mathrm{mM}]$& $A_i [\mathrm{mM}]$ & 
$L_i [\mathrm{mM}]$ & $M_i [10^{-5} \mathrm{mM}]$ &
$A_f [\mathrm{mM}]$ & $L_f [\mathrm{mM}]$ & $M_f [10^{-5} \mathrm{mM}]$ \\ \hline
$\bar{L}_{e1}=0.0277$ & $0.0969$ & $0.128$ & $7.521$ & $0.00398$ & $0.0937$ & $0.212$\\
$\bar{L}_{e2}=0.0278$ & $0.104 $ & $ 0.129$ & $8.600 $ & $0.00400 $ & $ 0.0940$ &  $0.212$\\
$\bar{L}_{e3}=0.0279$ & $0.109$ &$0.129$ & $9.406$ & $0.00400$ & $0.0946$ & $0.212$\\
$\bar{L}_{e4}=0.0618$ & $0.0141$ &$0.224$ & $0.360$ & - & - & - \\
$\bar{L}_{e5}=0.0620$ & $0.0141$ &$0.224$ & $0.360$ & $0.393$ & $0.285$ & $80.8$\\
$\bar{L}_{e6}=0.0630$ & $0.0141$ & $0.224$ & $0.360$ & $0.399$ & $0.289$ & $82.5$\\
$\bar{L}_{e7}=0.0650$ & $0.0141$ &$0.224$ & $0.360$ & $0.412$ & $0.298$ & $85.9$\\
\hline
\end{tabular}
\end{center}\end{small}
\vspace*{0.5cm}
\caption{\small \label{tab:points} Initial ($i$ subscript) and final ($f$ subscript) concentrations of allolactose ($A$), lactose ($L$), and mRNA ($M$) for simulations described in Section~\ref{sec:results}.}
\end{table}


\bibliographystyle{elsart-harv}

\bibliography{ecoli_jtb}

\begin{thebibliography}{25}
\expandafter\ifx\csname natexlab\endcsname\relax\def\natexlab#1{#1}\fi
\expandafter\ifx\csname url\endcsname\relax
  \def\url#1{\texttt{#1}}\fi
\expandafter\ifx\csname urlprefix\endcsname\relax\def\urlprefix{URL }\fi

\bibitem[{Alon et~al.(1999)Alon, Surette, Barkai, and Leibler}]{Alon}
Alon, U., Surette, M.~G., Barkai, N., Leibler, S., 1999. Robustness in
  bacterial chemotaxis. Nature 397, 168.

\bibitem[{Austin et~al.(2006)Austin, Allen, McCollum, Dar, Wilgus, Sayler,
  Samatova, Cox, and Simpson}]{Austin}
Austin, D.~W., Allen, M.~S., McCollum, J.~M., Dar, R.~D., Wilgus, J.~R.,
  Sayler, G.~S., Samatova, N.~F., Cox, C.~D., Simpson, M.~L., 2006. Gene
  network shaping of inherent noise spectra. Nature 439~(2).

\bibitem[{Beckwith(1987)}]{Beckwidth}
Beckwith, J., 1987. The lactose operon. In: Neidhardt, F.~C., Ingraham, J.~L.,
  Low, K.~B., Magasanik, B., Umbarger, H.~E. (Eds.), Escherichia coli and
  Salmonella: Cellular and Molecular Biology, Vol. 2. American Society for
  Microbiology, Washington, DC, pp. 1444--1452.

\bibitem[{Becskei and Serrano(2000)}]{Becskei}
Becskei, A., Serrano, L., 2000. Engineering stability in gene networks by
  autoregulation. Nature 405, 590.

\bibitem[{Casadesus and D'Ari(2002)}]{Casadesus_DAri}
Casadesus, J., D'Ari, R., 2002. Memory in bacteria and phage. Bioessays 24~(6),
  512--518.

\bibitem[{Cohn and Horibata(1959)}]{Cohn_Horibata}
Cohn, M., Horibata, K., 1959. Analysis of the differentiation and of the
  heterogeneity within a population of escherichia coli. Journal of
  Bacteriology 78, 613--623.

\bibitem[{DiLuzio et~al.(2005)DiLuzio, Turner, Mayer, Garstecki, Weibel, Berg,
  and Whitesides}]{DiLuzio}
DiLuzio, W., Turner, L., Mayer, M., Garstecki, P., Weibel, D.~B., Berg, H.,
  Whitesides, G., 2005. Eschericha coli swim on the right-hand side. Nature
  435~(30), 1271.

\bibitem[{Elowitz et~al.(2002)Elowitz, Levine, Siggia, and Swain}]{Elowitz}
Elowitz, M., Levine, A., Siggia, E., Swain, P., 2002. Stochastic gene
  expression in a single cell. Science 297~(5584), 1183--1186.

\bibitem[{Ferrell(2002)}]{Ferrell}
Ferrell, J.~E., 2002. Self-perpetuating states in signal transduction: positive
  feedback, double-negative feedback and bistability. Curr. Opin. Chem. Biol.
  6, 140--148.

\bibitem[{Gardiner(2004)}]{Gardiner}
Gardiner, C.~W., 2004. Handbook of Stochastic Methods. Springer, Berlin.

\bibitem[{Horsthemke and Lefever(1984)}]{Horsthemke_Lefever}
Horsthemke, W., Lefever, R., 1984. Noise-Induced Transitions. Springer-Verlag,
  Berlin.

\bibitem[{Laurent and Kellershohn(1999)}]{Laurent_Kellershohn}
Laurent, M., Kellershohn, N., 1999. Multistability: a major means of
  differentiation and evolution in biological systems. Trends Biochem. Sci.
  24~(11), 418--422.

\bibitem[{Mannella(2002)}]{Mannella}
Mannella, R., 2002. Eschericha coli swim on the right-hand side. Int. J. Mod.
  Phys. C 13~(9), 1177 -- 1194.

\bibitem[{Mathews et~al.(1999)Mathews, van Holde, and
  Ahern}]{Mathews_vHolde_Ahern}
Mathews, C.~K., van Holde, K.~E., Ahern, K.~G., 1999. Biochemistry. Prentice
  Hall.

\bibitem[{Monod et~al.(1952)Monod, Pappenheimer~Jr., and
  Cohen-Bazire}]{Monod_Pappenheimer}
Monod, J., Pappenheimer~Jr., A.~M., Cohen-Bazire, G., 1952. The kinetics of the
  biosynthesis of beta-galactosidase in escherichia coli as a function of
  growth. Biochim. Biophys. Acta 9~(6), 648.

\bibitem[{Novick and Weiner(1957)}]{Novick_Weiner}
Novick, A., Weiner, M., 1957. Enzyme induction as an all-or-none phenomenon.
  Proc. Natl. Acad. Sci. US A 43~(7), 553--568.

\bibitem[{Ozbudak et~al.(2004)Ozbudak, Thattai, Lim, Shraiman, and van
  Oudenaarden}]{Ozbudak}
Ozbudak, E.~M., Thattai, M., Lim, H., Shraiman, B., van Oudenaarden, A., 2004.
  Multistability in the lactose utilization network of escherichia coli. Nature
  427, 740.

\bibitem[{Paulsson(2005)}]{Paulsson}
Paulsson, J., 2005. Models of stochastic gene expression. Physics of Life
  Reviews 2, 157--175.

\bibitem[{Press et~al.(1993)Press, Flannery, Teukolsky, and Vetterling}]{nr}
Press, W., Flannery, B., Teukolsky, S., Vetterling, W., 1993. Numerical Recipes
  in C. Cambridge University Press, Cambridge.

\bibitem[{Rosenfeld et~al.(2005)Rosenfeld, Young, Alon, Swain, and
  Elowitz}]{Rosenfeld}
Rosenfeld, N., Young, J.~W., Alon, U., Swain, P.~S., Elowitz, M.~B., 2005. Gene
  regulation at the single-cell level. Science 307, 1962.

\bibitem[{Swain et~al.(2002)Swain, Elowitz, and Siggia}]{Swain_Elowitz}
Swain, P.~S., Elowitz, M.~B., Siggia, E.~D., 2002. Intrinsic and extrinsic
  contributions to stochasticity in gene expression. Proc. Natl. Acad. Sci. 99,
  12795--12800.

\bibitem[{Tabaka et~al.(2008)Tabaka, Cybulski, and Ho{\l}yst}]{Tabaka}
Tabaka, M., Cybulski, O., Ho{\l}yst, R., 2008. Accurate genetic switch in
  \textit{Escherichia coli}: Novel mechanism of regulation by co-repressor. J.
  Mol. Bio. 377, 1002--1014.

\bibitem[{Tsimring et~al.(2006)Tsimring, Volfson, and Hasty}]{Tsimring}
Tsimring, L.~S., Volfson, D., Hasty, J., 2006. Stochastically driven genetic
  circuits. Chaos 16, 026103.

\bibitem[{Yagil and Yagil(1971)}]{Yagil}
Yagil, G., Yagil, E., 1971. On the relation between effector concentration and
  the rate of induced enzyme synthesis. Biophys. J. 11, 11--27.

\bibitem[{Yildirim and Mackey(2003)}]{Yildirim_Mackey}
Yildirim, N., Mackey, M., 2003. Feedback regulation in the lactose operon: A
  mathematical modeling study and comparison with experimental data. Biophys.
  J. 84, 2841--2851.

\end{thebibliography}

\end{document}